\documentclass[%
reprint,
superscriptaddress,
 amsmath,amssymb,
 aps,
prb,
]{revtex4-2}

\usepackage{comment} 
\usepackage{graphicx}
\usepackage{epstopdf}
\usepackage{textcomp}
\usepackage{dcolumn}
\usepackage{bm}
\usepackage{gensymb}
\usepackage{upgreek}
\usepackage[usenames, dvipsnames]{color}
\usepackage{mathtools}
\usepackage{amsmath}
\usepackage{physics}
\usepackage{esvect}
\usepackage[mathscr]{euscript}
\usepackage{amssymb}
\DeclareUnicodeCharacter{2212}{-}
\DeclareUnicodeCharacter{03B4}{δ}
\DeclareUnicodeCharacter{2082}{₂}

\newcommand{\BaNi}{BaNi$_2$As$_2$}
\newcommand{\SrNi}{SrNi$_2$As$_2$}
\newcommand{\BaFe}{BaFe$_2$As$_2$}
\newcommand{\BaSr}{Ba$_{1-x}$Sr$_{x}$Ni$_2$As$_2$}

\newcommand{\BaSrLateTwo}{Ba$_{0.40}$Sr$_{0.60}$Ni$_2$As$_2$}
\newcommand{\Tc}{$T_c$}

\begin{document}

\preprint{APS/123-QED}

\title{Charge Order Evolution of Superconducting BaNi$_2$As$_2$ Under High Pressure}

\author{John Collini}
\affiliation{Maryland Quantum Materials Center, Department of Physics, University of Maryland, College Park, Maryland 20742, USA}

\author{Daniel J. Campbell}
\affiliation{Maryland Quantum Materials Center, Department of Physics, University of Maryland, College Park, Maryland 20742, USA}
\affiliation{Lawrence Livermore National Laboratory, Livermore, CA 94550, USA}

\author{Daniel Sneed}
\affiliation{Lawrence Livermore National Laboratory, Livermore, CA 94550, USA}

\author{Prathum Saraf}
\affiliation{Maryland Quantum Materials Center, Department of Physics, University of Maryland, College Park, Maryland 20742, USA}

\author{Christopher Eckberg}
\affiliation{Maryland Quantum Materials Center, Department of Physics, University of Maryland, College Park, Maryland 20742, USA}

\author{Jason Jeffries}
\affiliation{Lawrence Livermore National Laboratory, Livermore, CA 94550, USA}

\author{Nicholas Butch}
\affiliation{Maryland Quantum Materials Center, Department of Physics, University of Maryland, College Park, Maryland 20742, USA}
\affiliation{NIST Center for Neutron Research, National Institute of Standards and Technology, Gaithersburg, MD 20899, USA.}

\author{Johnpierre Paglione}
\email{paglione@umd.edu}
\affiliation{Maryland Quantum Materials Center, Department of Physics, University of Maryland, College Park, Maryland 20742, USA}
\affiliation{Canadian Institute for Advanced Research, Toronto, Ontario M5G 1Z8, Canada}

\date{\today}

\begin{abstract}
\BaNi, a non-magnetic superconductor counterpart to \BaFe, has been shown to develop nematic order, multiple charge orders, and a dramatic six-fold enhancement of superconductivity via isovalent chemical substitution of Sr for Ba. Here we present high pressure single-crystal and powder x-ray diffraction measurements of \BaNi~to study the effects of tuning lattice density on the evolution of charge order in this system. Single-crystal X-ray experiments track the evolution of the incommensurate ($Q$=0.28) and commensurate ($Q$=0.33 and $Q$=0.5) charge orders, and the tetragonal-triclinic distortion as a function of temperature up to pressures of 10.4~GPa, and powder diffraction experiments at 300 K provide lattice parameters up to 17~GPa. We find that applying pressure to \BaNi~ produces a similar evolution of structural and charge-ordered phases as found as a function of chemical pressure in \BaSr~, with coexisting commensurate charge orders appearing on increasing pressure. These phases also exhibit a similar abrupt cutoff at a critical pressure of ($9\pm 0.5$)~GPa, where powder diffraction experiments indicate a collapse of the tetragonal structure at higher temperatures. We discuss the relationship between this collapsed tetragonal phase and the discontinuous phase boundary observed at the optimal substitution value for superconductivity in \BaSr.  

\end{abstract}

\maketitle

Understanding the role of competing ground states in unconventional superconductors remains an important unresolved problem in quantum materials. In the cuprate \cite{CDW_Cuprate_Discovery_first,CDW_Cuprate_Discovery_second,CDW_Cuprate_Discovery_third,CDW_Cuprate_Discovery_fourth,Cuprates_Nature_Kiemer,Cuprates_Nematic_Auvray,Cuprates_Nematic_Kivelson,Cuprates_Nematic_Liu,Cuprates_Science_Orenstein} and iron-pnictide \cite{Fe-Sc_AdvPhys_Johnston,Fe-Sc_Nature_Paglione,Fe-Sc_RevModPhys_Stewart,Fernandes_Numatic_Review,Fisher_FeNumatic} superconducting families, antiferromagnetic fluctuations are thought to drive high temperature superconductivity, but coexisting and competing structural, charge and electronic nematic ordered phases have complicated any simple interpretation of the origin of high critical temperatures found in both systems. To explore the effectiveness of the other contributing phases in these systems, attention has turned towards studying related systems that show superconductive enhancement without a strong magnetic component. The nickel-based pnictide family, including \BaNi~and related compounds, has come into focus as an ideal platform for such investigation
\cite{Collini_BaSr_NoICCDW, Eckberg_nematic,Lee_BANI_CDW,Lee_Paper2,BaNi2As2_New_Pavlov,BaNi2As2_New_Rokharel,BaNi2As2_New_Yao,BaNi2As2Pressure_Park}. 

\BaNi~is considered a conventional fully-gapped superconductor with a transition temperature $T_c$ of 0.7 K, based on thermal conductivity measurements in the superconducting phase \cite{Kurita_BaNiThermal_SC}.
However, alkali earth substitution in the   \BaSr~series has recently been shown to reveal a rich phase diagram with coexisting superconducting, nematic, charge and structural orders \cite{Eckberg_nematic}, with provocative similarities to the other high-temperature superconductor families. 
\BaNi~is tetragonal and isostructural to its iron-based counterpart, \BaFe, at room temperature and ambient pressure. However, while \BaFe~undergoes a mild orthorhombic distortion at 140~K \cite{Rotter_BaFe2As2}, \BaNi~experiences a more severe first-order transition to a lower-symmetry triclinic structure at $T_s=135$ K \cite{Kothapalli_BANI_Neutron}. The tetragonal and triclinic phases of \BaNi~are characterized by distinct sets of Bragg peaks indexed to the space groups of $I4/mmm$ and $P1$ respectively. Here we use $(H,K,L)_{tet}$ and $(H,K,L)_{tri}$ separately to describe positions in momentum space for each phase. Further, unlike \BaFe, neutron measurements of \BaNi~have shown no evidence of magnetic order in the low temperature triclinic phase or anywhere else at ambient pressure \cite{Kothapalli_BANI_Neutron}.    
In this work, we apply pressure to study the evolution of structural and charge-ordered phases in \BaNi, comparing to those tuned by chemical pressure in \BaSr, to better understand the role of lattice density and structural bonding in stabilizing the rich phase diagram of the \BaNi~system.

\begin{figure}
    \centering
    \includegraphics[width=0.47\textwidth]{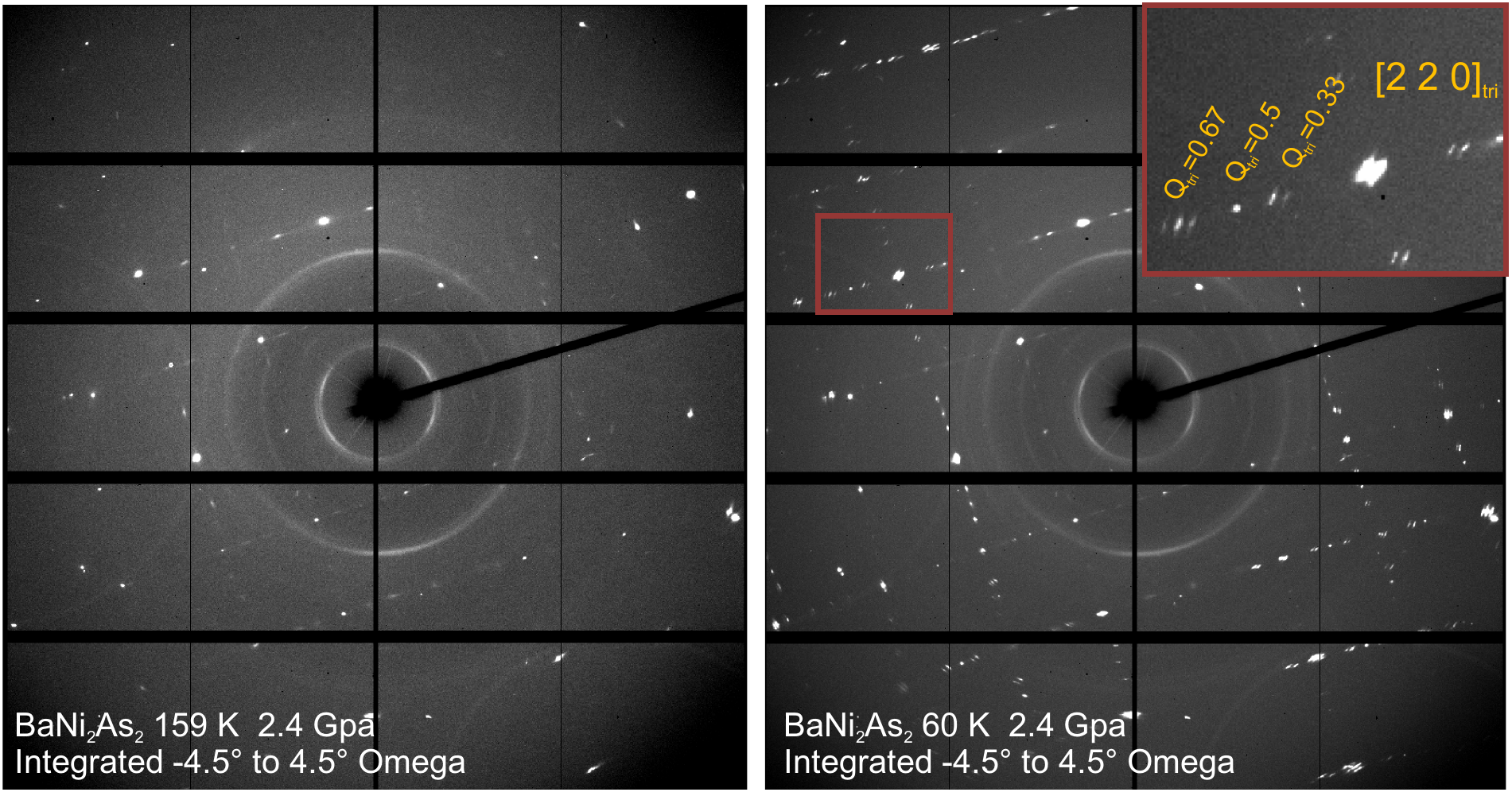}
    \caption{Single crystal x-ray diffraction images of \BaNi~ at 2.4~GPa taken at (a) 159~K and (b) 60~K. Highlighted in (b) is the [2 2 0]$_{tet}$ zone. Comparing high to low temperatures, we find the introduction of satellite peaks through-out the diffraction pattern. Superlattice peaks found midway between zones are evidence for the $Q$=0.5 C-CDW2 and superlattice peaks found at $1/3$ and $2/3$ way to the next zone are evidence for the $Q$=0.33 C-CDW1. Each image was exposed while continuously rocking the sample from -4.5\degree~to +4.5\degree~Omega.}
    \label{fig:Figure1_detector}
\end{figure}

\begin{figure*} 
    \centering
    \includegraphics[width=1\textwidth]
    {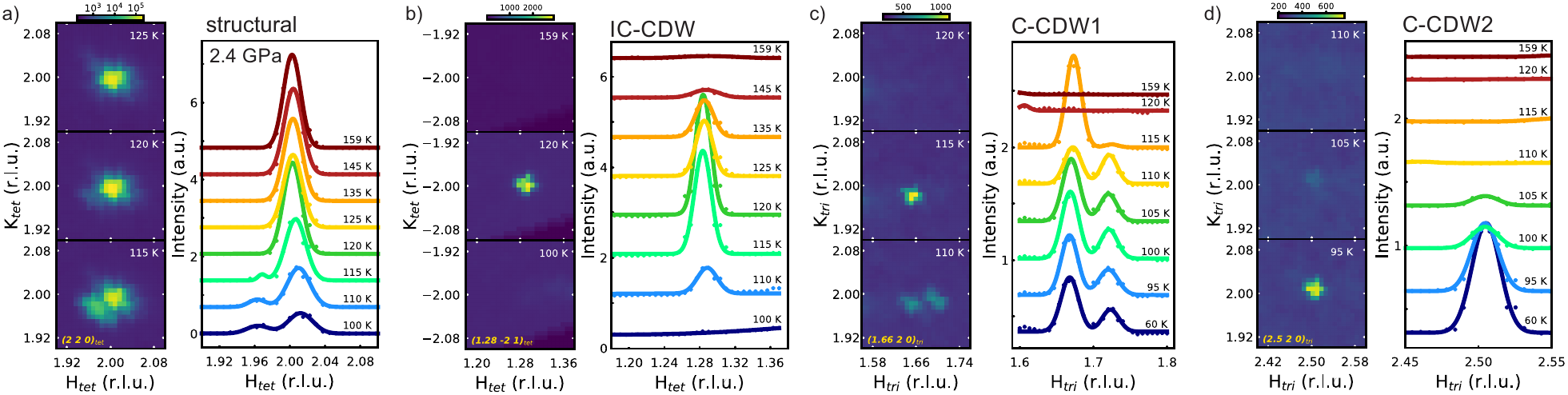}
    \caption{Four distinct features of \BaNi~at 2.4~GPa (Triclinic onset, IC-CDW, C-CDW1, and C-CDW2) shown as a function of decreasing temperature via momentum space plots and line cuts at a) [2 2 0]$_{tet}$, b) [1.28 -2 1]$_{tet}$, c) [1.66 2 0]$_{tri}$, and d) [2.5 2 0]$_{tri}$. We find the IC-CDW onsets at a temperature of 145 K and a vanishes at 105 K. We find the triclinic order, C-CDW1, and C-CDW2 onset at temperature of 120~K, 110~K, and 105~K respectfully. Line cuts are presented with Gaussian fits at the peaks for guidance.}
    \label{fig:Figure2_P2.4}
\end{figure*}

Recent X-ray diffraction experiments on the \BaSr~series have revealed the presence of several charge-ordered phases at ambient pressure and low temperatures. In \BaNi, a bi-directional incommensurate charge density wave (IC-CDW) develops in the tetragonal phase just above $T_s$ at $T_{IC}=148$~K and at a wavevector of $Q_{tet}=(0.28,0.28,0)_{tet}$ in a ``$4Q$'' state in the $ab$ plane \cite{Lee_Paper2}. At $T_s$, the incommensurate CDW vanishes and gives way to a commensurate CDW (C-CDW1) at wavevector $Q_{tri}=(0.33,0,0)_{tri}$ in the triclinic phase \cite{Lee_BANI_CDW}. As a function of Sr substitution ($x$), the IC-CDW onset falls closer in temperature to the triclinic onset before disappearing near $x$=0.6 \cite{Collini_BaSr_NoICCDW}. Additionally near $x$=0.4, a new commensurate charge order, C-CDW2, begins to form about 20~K below the triclinic distortion at $Q_{tri}=(0.5,0,0)_{tri}$, co-existing with the previously mentioned C-CDW1 down to base temperature with C-CDW2 being largely dominant in scattering intensity \cite{Lee_Paper2}. C-CDW2 eventually replaces C-CDW1 near $x$=0.6. At $x$=0.71, the triclinic and C-CDW2 phases abruptly vanish leaving only the tetragonal phase to exist down to base temperature up to $x$=1 \cite{Lee_Paper2}. Concurrently at the ``optimal'' Sr concentration of $x$=0.71, $T_c$ experiences a large six-fold enhancement up to a maximum of 3.5~K \cite{Eckberg_nematic}.

Elasto-resistance measurements of the $B_{1g}$ channel in \BaSr, corresponding to the symmetry-breaking strain along the [100] and [010] tetragonal crystallographic directions, have revealed large nematic susceptibilities and a nematic ordered phase that persists through a wide range of Sr substitution at temperatures above the triclinic distortion \cite{Eckberg_nematic}. At lower Sr concentrations, the nematic order co-exists with, and correlates strongly with, the IC-CDW intensity and crystallographic direction 
\cite{Eckberg_nematic,Lee_Paper2}. For $x$=0.65 and higher, the IC-CDW and nematic ordered phases both vanish \cite{Collini_BaSr_NoICCDW}, revealing an enhanced superconducting phase along with a strong nematic susceptibility above it \cite{Eckberg_nematic}. The correlations in intensity and co-existence for the IC-CDW and nematic phase suggest that the two phases are intimately related. 
With such sensitivity to nominally isovalent substitution, natural questions arise over the role of lattice density versus the impact of chemical substitution on the ordered phases in this system, and whether the evolution of these phases progresses differently for each tuning parameter.

Single crystal x-ray diffraction measurements under high pressure were taken at the 16-BMD beamline of the Advanced Phonton Source (APS) at Argonne National Laboratory (Figure \ref{fig:Figure1_detector}). Single crystals of \BaNi~were prepared in collaboration with the Equilibrium Physics at Extreme Conditions (EPEC) group of Lawrence Livermore National Laboratory using a diamond anvil cell (DAC) with a diamond culet size of 500~\micro m. The samples of \BaNi~were mounted with their c-axis normal to the diamond's culet, parallel to the x-ray beam. Neon gas was used as a pressure medium and ruby florescence was used as a pressure manometer \cite{Ruby_pressure}. A  200~$\mu$m diameter hole was drilled in the center of a stainless steel gasket. The gasket hole was filled with the sample, gas, and ruby, and closed at an intial pressure of 2.4~GPa. A beamline mounted cryostat controlled the temperature of the DAC and had working base temperature of about 37~K. A gas membrane was used to control the pressure remotely from outside cryostat and the hutch. The workable rocking angle for the experiment (denoted Omega on 16-BMD) was limited by the gasket hole to a range of -4.5~degrees to 4.5~degrees. Data were taken only on cooling for pressures 2.4~GPa, 5.4~GPa, 8.1~GPa, 9.5~GPA, and 10.5~GPa. Additionally, room temperature powder X-ray diffraction measurements were taken at beamline 16-IDB of the APS. A rhenium gasket was indented to about 40 um using 300 um culets, then a 100 um hole was drilled in the center. A powder of crushed \BaNi~crystals was loaded along with Cu powder and a ruby sphere. The ruby was used for initial offline pressure calibration \cite{BaFe2As2_SpinState} and pressure during the measurement was ultimately determined from the [200] reflection of Cu using the the equation of state \cite{Equation_of_State}. Mineral oil was used as the pressure medium, and a gas membrane was used to control pressure remotely from outside the hutch. High pressure electrical transport measurements were also performed up to 2.55~GPa using a piston-operated BeCu high pressure cell with Arcros Organics Perfluoro-compound FC-770\texttrademark~acting as the pressure medium.

\begin{figure} [!h!t]
    \centering
    \includegraphics[width=0.47\textwidth]{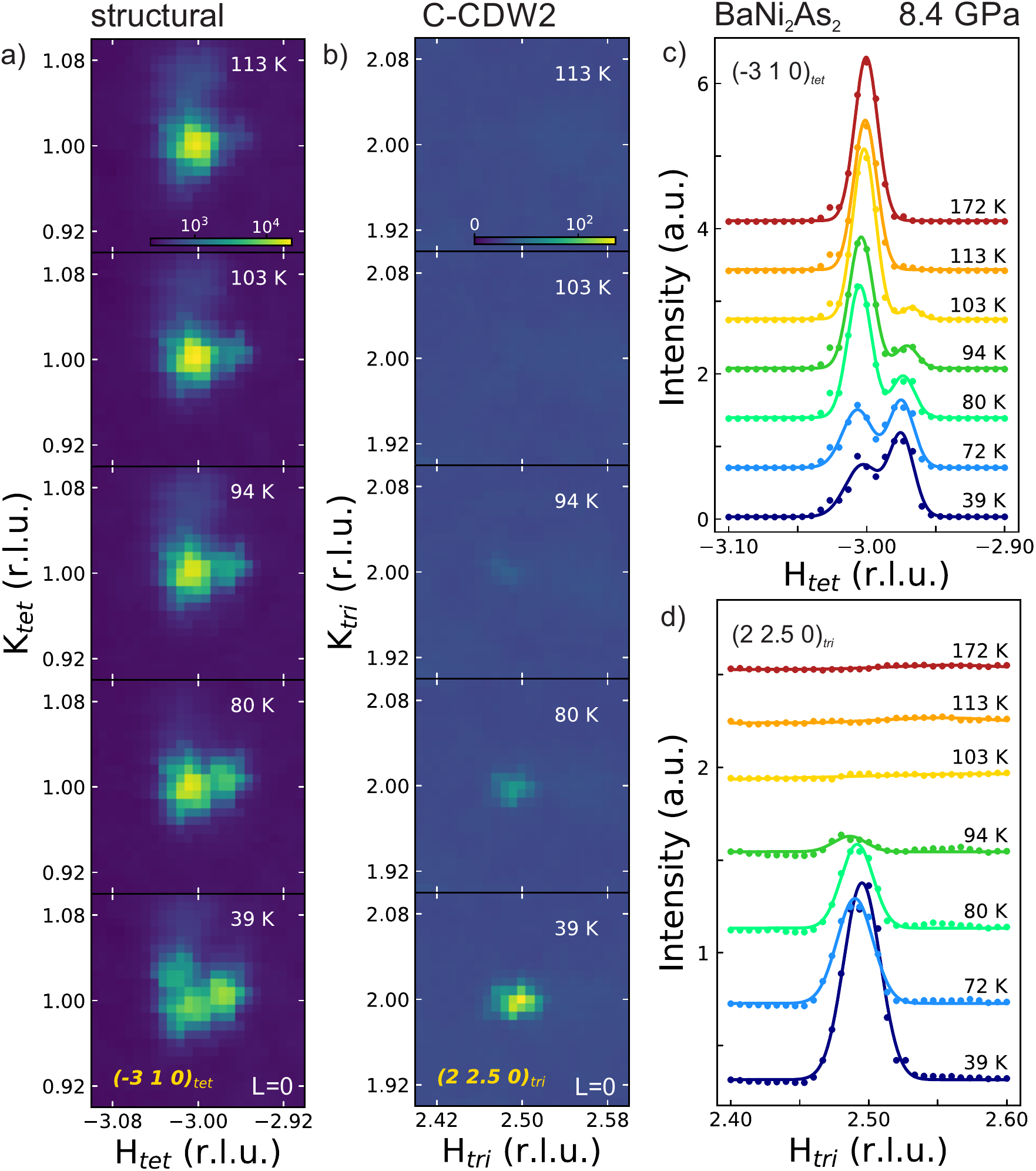}
    \caption{8.4~GPa x-ray data and linescans of a) C-CDW2 and b) triclinic order as seen by the [-3 1 0]$_{tet}$ and [2 2.5 0]$_{tri}$ peaks. We find that both phases onset at 103~K.}
    \label{fig:2.1}
\end{figure}

We find that charge order under pressure in \BaNi~develops in similar ways to Sr substitution, but with key notable differences. Starting at 2.4~GPa, we observe all three charge orders (C-CDW1, C-CDW2, and IC-CDW) and the triclinic distortion seen previously in the \BaSr~series (Figure \ref{fig:Figure2_P2.4}). We find the IC-CDW onsets at 140~K, which is 20~K above the triclinic distortion observed at 120~K at this pressure. The IC-CDW then vanishes at 105 K in the triclinic phase. This marks a different behavior from \BaSr, where we would instead expect the IC-CDW to vanish very close to the triclinic onset while still in the tetragonal phase \cite{Lee_Paper2}. For the commensurate charge order, C-CDW1 appears first at 110~K and C-CDW2 appears next at 105~K, both in the triclinic phase. C-CDW1 and C-CDW2 continue to coexist down to 60~K, the lowest temperature we measured for this pressure. This behavior, with the exception of a wide IC-CDW existence window in temperature, places \BaNi~at 2.4~GPa roughly in line with \BaSr~at a Sr\% between $x=$ 0.4 and 0.5.

As pressure is increased, the three charge orders and the triclinic distortion have their onset temperature decline and eventually vanish. At the next measured pressures of 5.3-5.6~GPa, C-CDW1 was not detected down to the lowest measured temperature of 49~K. C-CDW2 was first detected upon cooling at a temperature of 110 K for a pressure of 5.4 GPa. The triclinic phase onsets at 110 K for 5.4 GPa. At a higher pressure of 8.4 GPa, triclinic order and the C-CDW2 phase appear together at 103 K (Figure \ref{fig:2.1}), and at 9.5~GPa and 10.5~GPa, the system remains tetragonal down to at least 37~K, with no charge order or triclinic distortion detected.

 \begin{figure}
    \centering
    \includegraphics[width=0.47\textwidth]{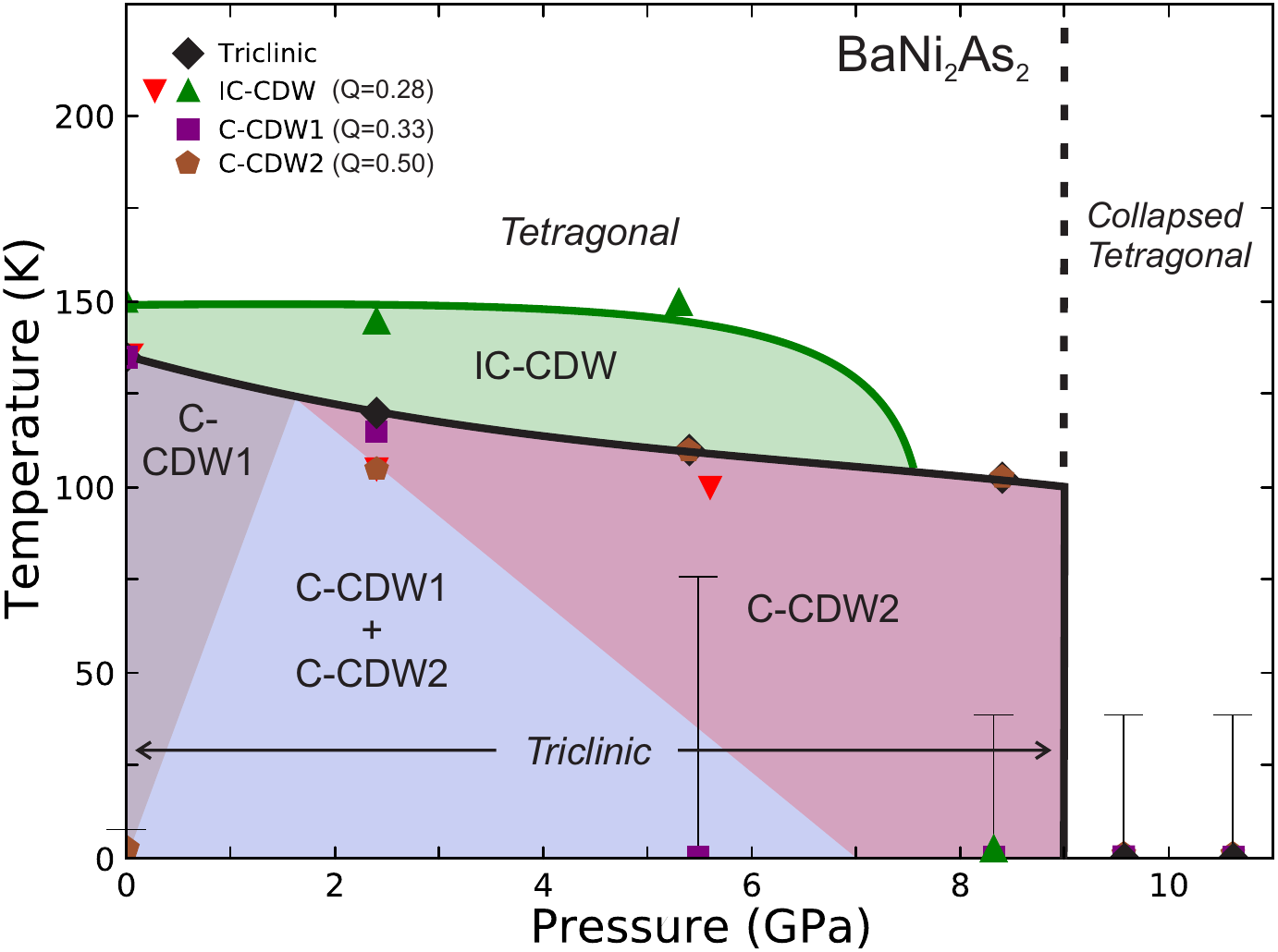}
    \caption{Pressure-temperature phase diagram of \BaNi~as determined by x-ray diffraction measurements. Highlighted are the IC-CDW, triclinic order, C-CDW1, C-CDW2, and collapsed tetragonal regions. The green triangles show the onset of the IC-CDW and the red triangles show the disappearance of the IC-CDW. The large error-bars seen at lower temperatures and higher pressures account for the absence of phases seen down to the lowest temperature measured for a given pressure.}
    \label{fig:3}
\end{figure}

The pressure phase diagram of \BaNi, presented in Figure 4, shows that the onset of all three of its charge-ordered phases and the triclinic distortion decline slightly in temperature as pressure is increased, before eventually vanishing separately. The IC-CDW onset holds roughly around 140~K with increasing pressure before vanishing beyond 5.4~GPa. The C-CDW1 onset likewise holds around 115~K before vanishing beyond 2.4~GPa. C-CDW2 and the triclinic distortion remain roughly in step with each other at 110~K through 100~K until their vanishings beyond 8.4~GPa. This picture compares well with the behavior seen in \BaSr~\cite{Lee_Paper2}.

\begin{figure} 
    \centering
    \includegraphics[width=0.47\textwidth]{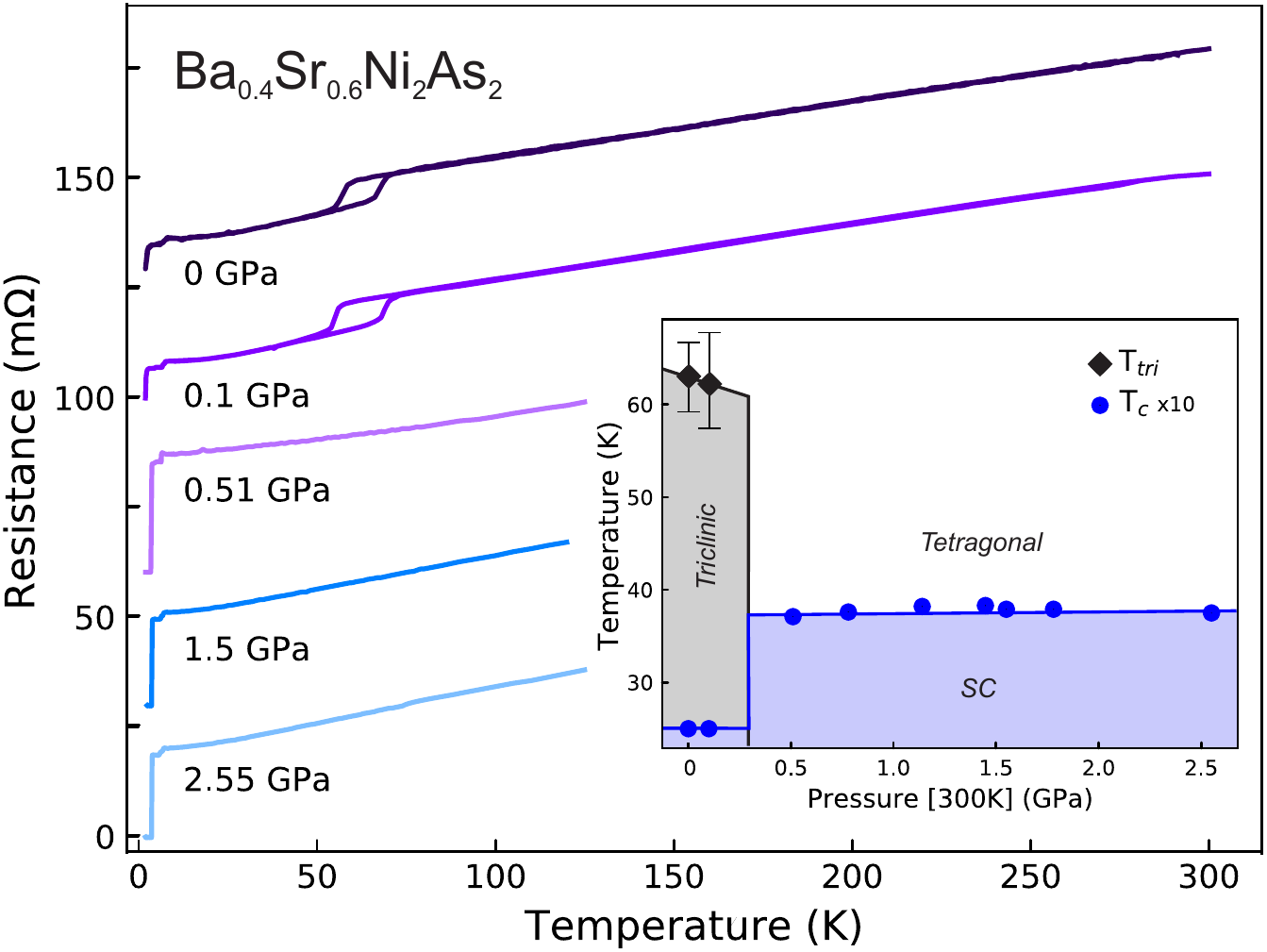}
    \caption{Resistance vs temperature of a \BaSrLateTwo~sample shown at selected pressures demonstrating the discontinuous phase boundary between the C-CDW2 + triclinic state and tetragonal state. Curves has been separated for readability. Lowest temperature reached for these measurements was 1.8~K. INSET: Pressure-temperature phase diagram of \BaSrLateTwo~taken from the resistivity data. \Tc~has been amplified by 10 for visibility.}
    \label{fig:4}
\end{figure}

We note that in \BaSr, the extinguished C-CDW2 and triclinic phases were observed to end discontinuously as a function of Sr concentration at $x=0.71$, before revealing a large enhancement in the bulk superconducting transition temperature \cite{Eckberg_nematic,Lee_Paper2}. In \BaNi~under applied pressure, we find a 8.4~GPa, 103~K C-CDW2 + triclinic state onset followed by a tetragonal-only state at 9.5~GPa down to 37~K. This strongly suggests a similar discontinuous extinction of the C-CDW2 and triclinic phases. Additional evidence supporting a discontinuous vanishing can be seen in a piston pressure cell transport measurement of \BaSrLateTwo~(Figure \ref{fig:4}). Between the narrowly separated pressures 0.1~GPa and 0.51~GPa, the sample of \BaSrLateTwo~goes from an ambient-pressure-like state of a C-CDW2 + triclinic phase with an onset temperature of ~60~K, to a complete vanishing of both phases followed by an enhanced superconducting temperature of 3.6~K. This enhanced \Tc~state persists through the highest recorded pressure of 2.55~GPa. The behavior induced in pressurized \BaSrLateTwo, coupled with our x-ray observations in pressurized \BaNi, and in previously reported ambient pressure \BaSr, point to there being a discontinuous phase boundary between triclinic + C-CDW2 and the enhanced superconducting state induced by a shrinking lattice caused by Sr isovalent substitution, pressure, or a combination of both.

\begin{figure} [!h!t]
    \centering
    \includegraphics[width=0.43\textwidth]{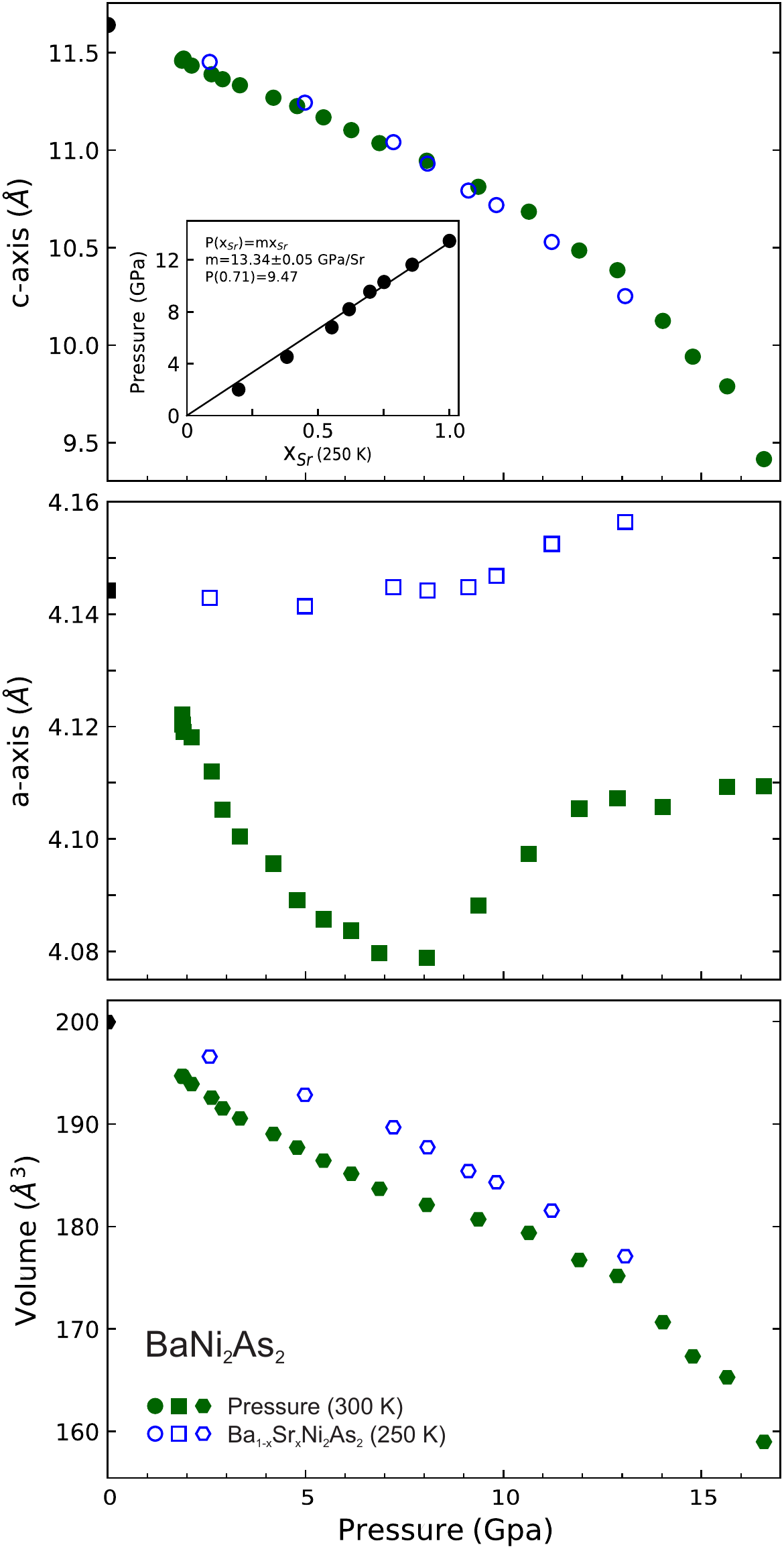}
    \caption{Lattice constant and unit cell volume data of 250~K \BaSr~(blue, open points) and 300~K pressurized \BaNi~(green, closed points) taken from powder x-ray data. The 250~K \BaSr~data were taken from the supplement of reference \cite{Eckberg_nematic}. The 300~K \BaNi~data were taken from pressurized powder x-ray diffraction data using the [110] peak for the a-axis and the [112] peak for the c-axis. The \BaSr~scale was chosen such that the c-axis of both systems align, visually showing the correlation between the two systems. INSET: points in pressure plotted against equivalent points in Sr substitution for alike c-axis values. A line, $P(x_{Sr})=mx_{Sr}$, is fit to the correlation, yielding $m=13.34\pm0.05$~GPa/Sr as the correlation constant relating the c-axis of pressure to the c-axis of \BaSr.}
    \label{fig:5}
\end{figure}

Lattice parameter studies for many systems with the ThCr$_2$Si$_2$-type ``122'' crystal structure have found a sensitive evolution of structure with pressure as a major driver of the interesting physics seen in their systems. In particular, a small-subset of this family exists very near a structural instability that leads to an isostructural collapse of the tetragonal unit cell. Called the collapsed tetragonal (CT) phase, this phase exhibits a severely reduced $c$-axis lattice parameter that can be driven by applied or chemical pressure, as shown in CaFe$_2$As$_2$ \cite{CT_CaFe2As2_RareEarth,CT_CaFe2As2_Pressure1,CT_CaFe2As2_Pressure2,CT_CaFe2As2_Pressure3,CT_CaFe2As2_Pressure4}, \BaFe~\cite{CT_BaFe2As2,CT_BaFe2As2_2}, SrFe$_2$As$_2$ \cite{CT_SrFe2as2}, CaRbFe$_4$As$_4$ \cite{CT_CaRbFe4As4}, LaRu$_2$P$_2$ \cite{CT_LaRu2P2}, and SrCo$_2$As$_2$ \cite{CT_SrCoAs}. 
In Figure \ref{fig:5}, we compare lattice parameters of \BaNi~up to 17~GPa pressures, measured at 300~K using powder x-ray diffraction, to those of \BaSr~measured at ambient pressure at 250~K \cite{Eckberg_nematic}. 
We first note the difference in temperatures between the two data sets of pressurized \BaNi~at 300~K and ambient-pressure \BaSr~at 250~K, but argue it is not significant enough to alter any meaningful comparisons. 
At 300~K, \BaNi~has reported $c$- and $a$-axis values of 11.6190~\AA~and of 4.1474 \AA, respectively \cite{BaNi2As2_lattice_numbers}, while \SrNi~has reported values of 10.290~\AA~and 4.154~\AA, respectfully \cite{SrNi2As2_lattice_numbers}. The 250~K \BaSr~lattice parameter dataset has values of 11.6423~\AA~and 4.1442~\AA~for x$_{Sr}$=0, as well as, 10.2515~\AA~and 4.1564~\AA~for x$_{Sr}$=1. The differences between 300~K and 250~K lattice parameters for \BaNi~and \SrNi~are fractions of a percent and therefore appropriate for our comparisons and discussions.

As shown in Fig.~6a), We find a correlation between the change in $c$-axis driven by Sr substitution and that driven by applied pressure in \BaNi, with both systems showing a monotonic decrease at a proportional rate. To directly compare, $c$-axis values of the pressurized \BaNi~dataset were linearly interpolated and matched with those of the $x_{Sr}$ series, fitting a linear relation $P(x_{Sr})=mx_{Sr}$ (Figure \ref{fig:5} insert) with $m$=$13.34\pm0.05$~GPa/Sr, between chemical pressure and physical pressure for \BaNi~along the c-axis. We test this correlation model with noteworthy points in x$_{Sr}$: the discontinuous phase boundary and maximum \Tc, $x_{C}=0.71$, as well as, the approximate ending point for the IC-CDW and nematic order, $x_{IC-CDW}=0.55$. Using this linear relationship, we find $P(x_{C})=9.47$~GPa and $P(x_{IC-CDW})=7.34$~GPa, closely predicting the corresponding events in the pressure.

In contrast, the evolution of the $a$-axis with pressure and Sr substitution (Fig.~6b) is not proportional. This is likely due to the difference between unconstrained chemical pressure, where Sr substition has a dominant effect in shrinking the $c$
-axis but nearly no effect on the in-plane bonding, while the quasi-hydrostatic pressure conditions in the DAC force both lattice constants to decrease in unison.
Interestingly, the nearly flat evolution of the $a$-axis in \BaSr~ shows a sudden increase near x$_{Sr}$=0.7, and likewise a more dramatic inflection occurs in pressurized \BaNi\ near $\sim$8~GPa, which are nearly equivalent effective pressure points according to the $c$-axis scaling above.
These both occur when the $c$-axis of both systems also shows a kink in near-linear evolution close to $\sim 10.8$~\AA. 

This peculiar evolution of $a$- and $c$-axis lattice constants is very similar to what is observed at the collapsed tetragonal (CT) phase transition of many Fe-pnictides \cite{CT_CaFe2As2_RareEarth,CT_CaFe2As2_Pressure1,CT_CaFe2As2_Pressure2,CT_CaFe2As2_Pressure3,CT_CaFe2As2_Pressure4,CT_BaFe2As2,CT_BaFe2As2_2,CT_SrFe2as2,CT_CaRbFe4As4}, where a bonding re-arrangement occurs when the As-As interlayer distance crosses about 3~\AA.
Generally, when either physical or chemical pressure is applied to these systems, there is a rise and fall in the $a$-axis with a valley to peak size ranging from 0.05~\AA~to 0.15~\AA, a discontinuous jump in the As-As interlayer distance, a flattening of the FeAs layer, and the most prominent change occurring in the $c$-axis lattice constant. This collapse is known to quench the Fe moment, suppressing its magnetism and having an effect on superconductivity \cite{CT_CaFe2as2_QuenchedMoment}. In pressurized \BaNi~and \BaSr, we find a similar CT transition developing via an $a$-axis expansion. Eckberg also noted that the rise in the $a$-axis coincides with d$_{As-As}$ crossing 3~\AA \cite{Eckberg_nematic}, the same threshold noted for the Fe-pnictide collapsed tetragonal phases. The onset of this CT transition in x$_{Sr}$ and pressure coincides with the charge order and triclinic order vanishing discontinuously. During this collapsed tetragonal transition, we speculate that the Ni-As layers begin to flatten along the $ab$ plane for the duration of this a-axis expansion, similarly to the flattening Fe-As layers in the CT phase of Fe-pnictides. We thus infer that the CT transition in \BaNi~plays a key role in abruptly extinguishing both charge order and the triclic structural phase. 
Open questions remain as to why the CT transition in \BaSr\ is much subtler than that found in iron-based materials and even that observed here in \BaNi\  under applied pressure. Further study into crystallographic parameters, Fermi surface and transport response and spectroscopies may help elucidate these questions.

In conclusion, we have tracked the pressure and temperature dependence of the charge orders, the triclinic distortion, and the lattice parameters of \BaNi~and have compared them to previously reported results in \BaSr. We find that the four phases, IC-CDW, C-CDW1, C-CDW2, and triclinic order slowly decline in onset temperature as a function of pressure until they eventually vanish at higher temperature. We also find that C-CDW2 and triclinic order, the final phases to survive, vanish discontinuously between 8.4~GPa and 9.5~GPa. At this point in pressure, we find a collapsed tetragonal transition in \BaNi~via an expanding a-axis as measured through powder x-ray diffraction. We speculate that this collapsed tetragonal phase plays a role in extinguishing charge order and enhancing superconductivity plus nematic susceptibility immediately following it. These data continue to present \BaNi~as a rich, non-magnetic superconducting system worthy of study.  

This work is supported by the U.S. National Science Foundation Grant No. DMR1905891, and the Gordon and Betty Moore Foundation’s EPiQS Initiative through Grant No. GBMF9071. Work at LLNL was supported by the U.S. Department of Energy, Office of Science,
Office of Workforce Development for Teachers and Scientists, Office of Science Graduate Student Research (SCGSR) program, which is administered by the Oak Ridge Institute for Science and Education (ORISE) for the DOE. ORISE is managed by ORAU under contract number DE‐SC0014664. All opinions expressed in this paper are of the authors and do not necessarily reflect the policies and views of DOE, ORAU, or ORISE. Experiments were also funded in part by a QuantEmX grant from the Institute for Complex Adaptive Matter and the Gordon and Betty Moore Foundation through Grant GBMF9616. Portions of this work were performed at HPCAT (Sector 16), Advanced Photon Source (APS), Argonne National Laboratory. HPCAT operations are supported by DOE-NNSA’s Office of Experimental Sciences. The Advanced Photon Source is a U.S. Department of Energy (DOE) Office of Science User Facility operated for the DOE Office of Science by Argonne National Laboratory under Contract No. DE-AC02-06CH11357. This work was performed under the auspices of the U.S. Department of Energy by Lawrence Livermore National Laboratory under Contract DE-AC52-07NA27344.

\bibliographystyle{apsrev4-2}
\bibliography{BSNA_CDW}

\end{document}